\documentclass{annphys_coll}
\usepackage{graphicx,amsmath}
\usepackage[latin1]{inputenc}
\usepackage[OT1]{fontenc}
\newcommand{\eg}{\emph{e.g.}}
\begin{document}

%%-----------------------------
%%      the top matter
%%-----------------------------
\title{Wheeler's delayed-choice thought experiment: Experimental realization and theoretical analysis} 
\author{V. Jacques}%
\address{Laboratoire de Photonique Quantique et Moléculaire, UMR CNRS 8537, École Normale Supérieure de Cachan, France}
\author{E Wu}\sameaddress{1}\secondaddress{Key Laboratory of Optical and Magnetic Resonance Spectroscopy, East Chinan Normal University, Shanghai, Chine}
\author{F. Grosshans}\sameaddress{1}
\author{F. Treussart}\sameaddress{1}
\author{A. Aspect}\address{Laboratoire Charles Fabry de l'Institut d'Optique, UMR CNRS 8501, Palaiseau, France}
\author{Ph. Grangier}\sameaddress{3} 
\author{J.-F. Roch}\sameaddress{1}
\maketitle
\begin{abstract} 
Wheeler has strikingly illustrated the wave-particle duality by the 
delayed-choice thought experiment, in which the configuration of a 2-path
interferometer is chosen after a single-photon light-pulsed has entered it.
%:
%either the interferometer is closed (the two paths are recombined) and the 
%interference is observed, or it stays open and the path followed by the photon
%is observed. 
%We present an almost ideal experimental realization of Wheeler's
%experiment and a quantitative theoretical analysis of the experimental results.
%with single photons, allowing an unambiguous measurement of the
%followed path. The configuration choice (open or closed) is causally separated
%from the entrance of the photon into the interferometer.
We present a quantitative theoretical analysis of an experimental realization
of Wheeler's proposal.
\end{abstract}
%
%%-----------------------------
%%      your text
%%-----------------------------
\section{Introduction}

Wheeler has strikingly illustrated the wave-particle duality when he proposed
the delayed-choice thought-experiment, 
%where the configuration of a two-path  interferometer is changed
where the output beamsplitter of a Mach-Zehnder interferometer (MZI) is inserted or removed
after a single-photon light-pulse has entered it \cite{WheelerDCE}.
Since the wave-like 
or particle-like behavior of the light-pulse can be shown with these complementary experimental setups,
this wave-particle duality seems consistent with a light
choosing to be a wave or a particle, according to 
the experimental setup.
Wheeler's delayed-choice experiment challenges such a naive interpretation
by delaying the experimental setup choice, so that the light-pulse
can only ``learn'' this choice after having entered the setup.

%The wave-like behavior of the light is tested by a Mach-Zehnder closed interferometer,
%and its particle-like behavior by a Hanbury Brown and Twiss open interferometer.
%Both are two-path interferometers, and their only difference is the presence or
%not of the output beamsplitter. Wheeler has proposed to choose to insert or 
%remove this beamsplitter after the entrance of the photon into the 
%interferometer.
\section{Experimental realization}
\begin{figure}%[b]
\centerline{\includegraphics[height=2.5cm]{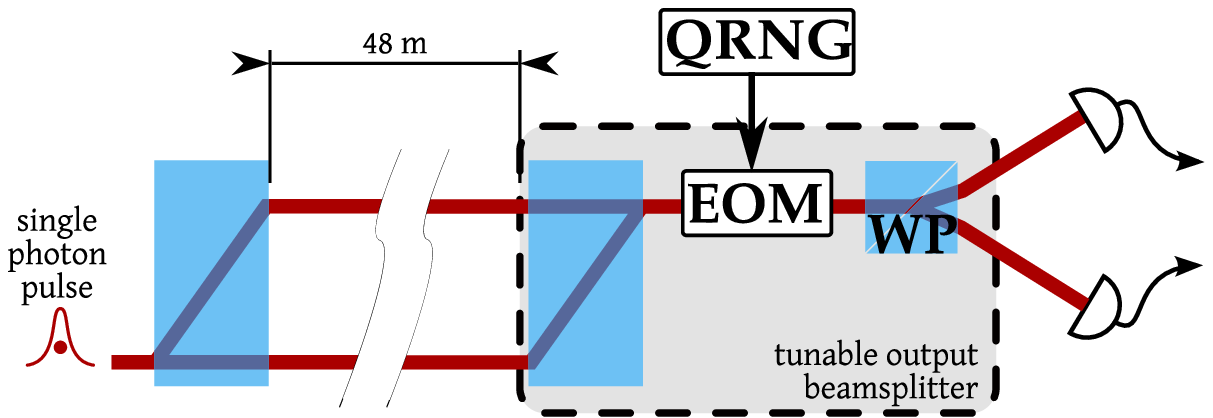} \includegraphics[height=4cm]{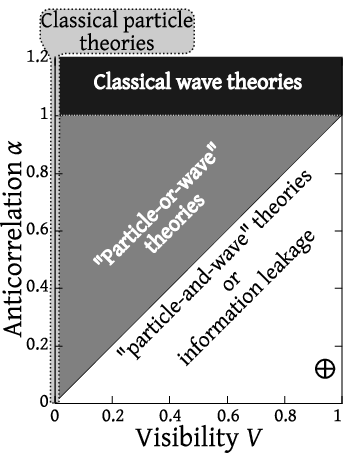}}
\caption{\textit{Left: sketch of the experimental setup. Right: possible theories compatible with the 
 parameters $V$ and $\alpha$. $\oplus$ corresponds to the experimental results.}}
\label{experiment}
\end{figure}

We recently realized almost ideally this thought experiment \cite{SciDCE}. 
The light pulses are true single-photons, emitted by a NV¯ color center in a  
diamond nanocrystal \cite{Alexnano}.
The photon number of the pulse is checked by observing an
anticorrelation parameter $\alpha=12\%<1$ 
on the output ports of a beamsplitter\cite{OldPhot}.

As shown in figure \ref{experiment},
we have built a Jamin polarization interferometer, followed by an 
electro-optical modulator (EOM) and a Wollaston prism (WP).
This interferometer is equivalent to a MZI 
%Mach-Zehnder interferometer 
where the
output beamsplitter reflection coefficient can be tuned from 1/2 (closed
configuration) to 0 (open configuration) in 40~ns. 
The two paths of this interferometer are spatially separated by 5~mm and propagate
over 48~m, which corresponds to a propagation time of 160~ns. 

The configuration choice is performed by a quantum random number generator 
(QRNG) placed at the interferometer output. 
The timing ensures that any information about the configuration
choice should travel at 4 times the speed of light to influence the photon 
behavior at the entrance \cite{SciDCE}.

We have observed interference with a visibility $V=94\%$ as well as an 
anticorrelation parameter $\alpha=12\%$, both in the delayed-choice mode, 
in full agreement with quantum mechanics predictions. 

%Nous avons également utilisé notre système dans des régime intermédiaire, où la
%complémentarité se manifeste par un compromis entre la propriété ondulatoire mise en évidence par l'inter-
%férence et les propriétés corpusculaires, mises en évidence sur l'information sur le chemin suivi.

\section{Theoretical analysis}
As stated above, the goal of the experiment is to disprove theories where the 
light-pulse behaves either like a wave or a particle, depending on the experimental setup.
We present here a quantitative model of this behavior,
and confront it to the experimental results reported above.
Note that such experiment can  be perfectly explained by other theories 
exhibiting a wave-particle duality, like the de~Broglie-Bohm pilot-wave 
theory \cite{BohmDCE, BellDCE}.
Wheeler's delayed-choice experiment 
%is not intended to test against 
%de~Broglie-Bohm 
%pilot-wave theory. 
%It 
tests against particle-OR-wave theories, as opposed to particle-AND-wave
theories, like quantum mechanics or the pilot-wave theory.

In a particle-or-wave theory, the behavior of the light-pulse  
is quantified by the probability $p_w$
to behave like a wave, and the complementary probability $p_p=1-p_w$ to behave like a 
particle. 
%The different behavior of the light-pulse depending 
The dependence of this behavior
on the interferometer
configuration is characterized by the probability $p_{w|o}$ (resp. $p_{w|c}$)  
%to behave like a wave 
of a wave-like behavior
when the 
interferometer is open (resp. closed).

When the interferometer configuration choice is random and relativistically
separated from the entrance of the photon in the interferometer, this choice
 cannot influence the photon behavior,
imposing $p_{w|o}=p_{w|c}$.

The interference visibility $V$ quantifies the wave aspect of the light-pulse
when the interferometer is closed.
It depends of the optical pathlength difference between the two arms, 
which can only influence 
something (\eg\ a wave) which simultaneously travels along bthe two arms. A 
single classical particle, on the contrary, has to choose its path 
and will not be sensitive to this path difference, leading to $V=0$
when the light-pulse behaves like a particle. 
Therefore, one has $V\le p_{w|c}$.

The anticorrelation parameter $\alpha$ 
%is greater than one 
obeys $\alpha\ge1$
for any theory where the light is described by a wave, but decreases
to zero for classical single particles.
It is therefore a measure of the 
particle-like behavior\cite{OldPhot},
with the constraint
$\alpha\ge p_{w|o}$.

These inequalities show that a wave-or-particle theory can only explain
experimental results in the delayed-choice regime if $V\le\alpha$.
%In our setup, the random number generator exits the past lightcone of the
%entrance of the photon in the interferometer 160~ns before the actual random number
%generation. $p$ should therefore be evaluated taking into account the possibility
%to guess the output of the RNG between 115 and 160~ns before it actually occurs. 
%The autocorrelation of the output at 240~ns being around 4\%, a 
%difference $|p_w-p_w'|$exceeding
%a few percents seems excluded.
%
%Note that our experimental setup has a very low efficiency,
%giving rise to approximately an event every 7000 clock cycle on average. 
%Due to this low efficiency, one should take into account the predictability 
%of the random number generator, conditioned on a positive detection
%event. Either one assumes the detection probability to be independent of the QRNG 
%(fair sampling assumption), or one need to perform a more detailed analysis of the %QRNG.
%We will assume fair sampling here, deferring the more detailed analysis of the QRNG for later. 
% 
%\subsection{Bounds on information from the experimental parameters}
%
On the other hand, 
if $V>\alpha$, as in our experiment where $V=94\%$ and $\alpha=12\%$,
this can only be explained by
either a wave-and-particle theory, like quantum mechanics, 
or by an information leakage from the QRNG to the light-pulse.

To quantify the amount of information leakage needed to explain the observed 
values of the parameters $V$ and $\alpha$, we need to compute a lower bound
for the probability of the light-pulse to correctly ``guess'' the interferometer 
configuration. 
This probability is constrained by the conditional
probability $p_{c|w}$ (resp. $p_{o|p}$) for the interferometer
to be closed (resp. open) knowing the light behaves like a wave (resp. like a particle).
Applying Bayes theorem, one has
\begin{align*}
p_{c|w}&=\frac{p_{w|c}}{p_{w|o}+p_{w|c}}\ge \frac{V}{V+\alpha}>88\%;
			&p_{o|p}&=\frac{1-p_{w|o}}{2-p_{w|o}-p_{w|c}}\ge\frac{1-\alpha}{2-V-\alpha}>93\%.
\end{align*}
Assuming a symmetrical probability for the light to correctly ``guess'' 
the QRNG output
(\ie\ $p_{c|w}=p_{o|p}$), this probability is therefore  
greater than 93\%, which is hardly compatible with the estimated 52\% predictability of the 
QRNG \cite{SciDCE}.

In conclusion, the experimental results can only be explained if the light-pulse is described by either
(1) a wave-and-particle theory, like quantum mechanics;
(2) a wave-or-particle theory where information on the setup can travel at 4 times the speed of light;
(3) a wave-or-particle theory and our QRNG exhibits undetected correlationstaht could be detected with probability a higher than 93\%. 
Needless to say, the most reasonable option seems to be the first one!

%Note that if $V\le\alpha$, the above equation are compatible with
%$p_{c|w}=p_{o|p}=\tfrac12$
%\ie the behavior of the photon can be independent from the experiment configuration.

%\section{conclusion}
%We have demonstrated an almost ideal realization of delayed-choice experiment. 
%We have also theoretically proved that ,
%short of a faster than light information transfer, 
%the obtained experimental results can only be explained by a dual wave-and-particle theory like quantum mechanics.
%%-----------------------------
%%      your bibliography
%%-----------------------------

\end{document}